\documentclass[prb,reprint,twocolumn,aps]{revtex4}
\usepackage{amsfonts}
\usepackage{graphicx, subfigure}
\usepackage{epsfig}
\usepackage{lscape}
\usepackage{rotating}
\begin{document}

\title{Ab initio van der Waals interactions in simulations of water alter structure from mainly tetrahedral to high-density-like}

\author{Andreas M\o gelh\o j$^{a,d}$, Andr\'{e} Kelkkanen$^{a,b}$, K. Thor Wikfeldt$^c$, Jakob Schi\o tz$^g$, Jens J\o rgen Mortensen$^a$, Lars G.M. Pettersson$^c$, Bengt I. Lundqvist$^{a,b}$, Karsten W. Jacobsen$^a$, Anders Nilsson$^d$ and Jens K. N{\o}rskov$^{a, e,f}$}

\affiliation{$^a$Center for Atomic-scale Materials Design (CAMD),Department of Physics, Building 307, Nano DTU, Technical University of Denmark, DK-2800 Kgs. Lyngby, Denmark}
\affiliation{$^b$Department of Applied Physics, Chalmers University of Technology SE-412 96 G\"oteborg, Sweden}
\affiliation{$^c$Department of Physics, AlbaNova, Stockholm University, SE-106 91 Stockholm, Sweden}
\affiliation{$^d$Stanford Synchrotron Radiation Lightsource, 2575 Sand Hill Road, Menlo Park, California 94025, USA}
\affiliation{$^e$SLAC National Accelerator Laboratory, 2575 Sand Hill Road, Menlo Park,  California 94025, USA}
\affiliation{$^f$Department of Chemical Engineering, Stanford University, Stanford, CA 94305, USA}
\affiliation{$^g$Center for Individual Nanoparticle Functionality (CINF), Department of Physics, Technical University of Denmark, DK-2800 Kongens Lyngby, Denmark}

\date{\today}

\begin{abstract}
The structure of liquid water at ambient conditions is studied in {\it ab initio} molecular dynamics simulations in the NVE ensemble using van der Waals (vdW) density-functional theory, {\it i.e.} using the new exchange-correlation functionals optPBE-vdW and vdW-DF2, where the latter has softer non-local correlation terms. Inclusion of the more isotropic vdW interactions counteracts highly directional hydrogen-bonds, which are enhanced by standard functionals. This brings about a softening of the microscopic structure of water, as seen from the broadening of angular distribution functions and, in particular, from the much lower and broader first peak in the oxygen-oxygen pair-correlation function (PCF) and loss of structure in the outer hydration shells. Inclusion of vdW interactions is shown to shift the balance of resulting structures from open tetrahedral to more close-packed. The resulting O-O PCF shows some resemblance with experiment for high-density water (A. K. Soper and M. A. Ricci, Phys. Rev. Lett., 84:2881, 2000), but not directly with experiment for ambient water. Considering the accuracy of the new functionals for interaction energies we investigate whether the simulation protocol could cause the deviation. An O-O PCF consisting of a linear combination of 70\% from vdW-DF2 and 30\% from low-density liquid water, as extrapolated from experiments, reproduces near-quantitatively the experimental O-O PCF for ambient water. This suggests the possibility that the new functionals may be reliable and that instead larger-scale simulations in the NPT ensemble, where the density is allowed to fluctuate in accordance with proposals for supercooled water, could resolve the apparent discrepancy with the measured PCF.
\end{abstract}

\maketitle

\section{Introduction}
Liquid water plays a crucial role in all biological and numerous chemical processes, which has provided the incentive for many detailed experimental and theoretical studies probing both structural and dynamical properties of the fluid. However, the microscopic structure of ambient liquid water is still a matter of current debate.~\cite{Myneni2002,Wernet2004, Smith2004,Sciencecomment, Hetenyi2004, Cavalleri2005,Kennedy2005, Odelius2006,Prendergast2006,Lee2006, leetmaa2008diffraction, fuchs2008isotope, tokushima2008high, XEScomment, Wikfeldt2009, PNAS2009,Mallamace2009, odelius2009molecular, PNAScomment, PNASreply, Chen2010,Clark2010,ClarkMolPhys2010,Soper2010,Tokushima2010,Nilsson2010,Leetmaa2010} In particular two classes of models are currently being considered, where the traditional model of water is based on a continuous distribution of distorted tetrahedral structures. This is typical of what most molecular dynamics simulations currently give. However, most of these simulations give over-structured O-O and O-H pair-correlation functions (PCFs) and show discrepancies in comparison to x-ray and neutron scattering experimental data.~\cite{leetmaa2008diffraction,Wikfeldt2009} It is, however, possible to generate a more distorted tetrahedral structure model that is consistent with the diffraction data, but equivalent agreement is seen also for alternative asymmetrical and mixture models illustrating that diffraction data do not discriminate between differently hydrogen-bonded (H-bonded) structure models.~\cite{Soper2005Asym, leetmaa2008diffraction,Wikfeldt2009} 

Based on recent findings correlating x-ray emission spectroscopy (XES) with x-ray Raman scattering (XRS) and small angle x-ray scattering (SAXS) data,~\cite{PNAS2009} a model has been suggested where a division into contributions from two classes of local instantaneous H-bonded structures is driven by incommensurate requirements for minimizing enthalpy and maximizing entropy; in particular the XES data show two well-separated peaks which interconvert but do not broaden with changes in temperature.~\cite{tokushima2008high,PNAS2009, PNASreply,Tokushima2010} In the proposed picture the dominating class at ambient temperatures consists of a continuum of structures with some resemblance to high-pressure water,~\cite{PNAS2009} but with a further expanded first shell (more distorted H-bonds) and more disorder in the 2nd shell; this was based on the temperature dependent shift of the dominating peak in the XES spectra indicating more thermal distortion and disorder with increasing temperature. The second class corresponds to fluctuations where regions of strongly tetrahedral structures (similar to low-density water) appear in different sizes and shapes as the molecules attempt to form enthalpically favored tetrahedral H-bond structures, resulting in mean size interpreted from the SAXS data as $\sim$1nm,~\cite{PNAS2009} but naturally many sizes and shapes would appear. It should be emphasized that since these are fluctuations no strict boundaries between the two classes should be expected.  

The attosecond (XRS, SAXS) to femtosecond (XES) time scales of the experimental probes are too fast for molecular motion to be followed and the experimental data thus correspond to a statistical sampling of instantaneous, frozen local structures in the liquid; no experimental information on the time scale of such fluctuations is thus currently available.~\cite{PNAS2009} Besides being consistent with both neutron and x-ray diffraction,~\cite{Wikfeldt2009} this picture was recently also shown to bring a consistency between x-ray diffraction and extended x-ray absorption fine structure (EXAFS) data, requiring both disordered structures and a fraction of molecules with straight, strong hydrogen-bonds.~\cite{Wikfeldt2010} Other opinions, however, exist regarding the interpretation of the new SAXS, XES and XRS data.~\cite{PNAScomment, odelius2009molecular, fuchs2008isotope, Smith2004,Prendergast2006, Chen2010, Clark2010, ClarkMolPhys2010,Soper2010} On the other hand, recent SAXS data extending into the supercooled regime and supported by theoretical simulations~\cite{Huang2010,Wikfeldt2011} as well as recent high-quality x-ray diffraction data resolving shell structure out to 12 {\AA} even in ambient and hot water, but contributed by a minority species,~\cite{Huang2011} provide support for the original interpretation in this debate.

Most simple rare gas solids and liquids have a nearest-neighbor coordination of 12 whereas hexagonal ice, due to the directional H-bonds has a coordination of only 4. The latter leads to large open volumes in the ice lattice and a resulting low density. The dispersion, or van der Waals (vdW) force, in condensed rare gases leads to non-directional, isotropic interactions and closer packing. Similarly, the inclusion of vdW interactions in {\it ab initio} simulations of water may counteract the directional interactions and lead to better agreement with, {\it e.g.}, experimental PCFs. Here it should be understood that, while this could be regarded as a minimum requirement of a water model, it is by no means sufficient for a complete description.  Interestingly, it has been argued on thermodynamic grounds that over a large range of the liquid-vapor coexistence line the averaged water interaction potential should resemble that of liquid Argon~\cite{Lishchuk2010}, {\it i.e.} not be determined by directional H-bonding.

Water shows many anomalies in its thermodynamic properties, such as compressibility, density variation and heat capacity.~\cite{debenedetti2003supercooled,Stanley2010,Angell1982} In attempts to explain this, directional H-bonds and more isotropic vdW forces are key concepts. While vdW forces are well defined as results of non-local electronic correlations, there is no unique way to characterize H-bonds in terms of topology or interaction strength. And yet "the H-bond governs the overall structure and the dynamics of water".~\cite{Mallamacekort2009} 

One of the models to explain the enhanced anomalies in supercooled water is the liquid-liquid critical-point (LLCP) hypothesis,~\cite{Speedy1982,Stanley1980,Poole1992} with the most substantial role played by cooperative H-bond interactions among the water molecules.~\cite{stokely2010cellmodel} The LLCP model explains the significant increase in density fluctuations upon supercooling water, which is evidenced by the anomalously increasing isothermal compressibility,~\cite{speedy1976isothermal} as resulting from attempts to locally form enthalpically favored open tetrahedrally coordinated H-bond regions. It furthermore connects the deeply supercooled liquid state of water to the polyamorphism seen in ices, {\it i.e.} the distinct low-density and high-density amorphous ice phases (LDA/HDA). A high-density liquid (HDL) phase transforms to an ordered low-density phase (LDL) in the deeply supercooled region through a first-order phase transition at high pressures above the LLCP and through a continuous smooth transition upon crossing the Widom line at pressures below the critical.~\cite{Stanley_review,Poole1994,Borick1995,Poole1992,Brovchenko2006} There are differences in their respective local structures; in pure HDL the local tetrahedrally coordinated H-bond structure is perturbed by a partially collapsed second coordination shell, while in the LDL a more open and locally "bulk-ice-like" H-bond network is realized.~\cite{soper2000highdens,Mallamace2009,Stanley_review} 

The combined XES, XAS and SAXS results described above,~\cite{PNAS2009} which indicate nanoscale density and structural fluctuations, can be easily interpreted as reflections of this "competition" between the two local forms, HDL (maximizing entropy) and LDL (minimizing enthalpy) and thus viewed as extending an established picture of supercooled water into the ambient regime. Whether HDL and LDL can exist as pure phases, accompanied by a liquid-liquid phase transition and a critical point, is still unresolved and alternative models, {\it e.g.}, singularity-free (SF),~\cite{Stanley1980, Sastry1996} critical-point-free (CPF)~\cite{Angell2008} and stability limit (SL) conjecture~\cite{Speedy1982SL} scenarios have been proposed, however still building on structural HDL/LDL fluctuations.

In the quantitative characterization of water, computer simulations play a vital role. Empirical force fields are frequently applied but with questionable transferability, since force fields are parameterized against experimental data or against a by necessity limited set of quantum chemically computed structures. Furthermore, many-body interactions beyond pair-interactions are frequently not taken into account. 

These deficiencies are eliminated in Car-Parrinello~\cite{Car-Parrinello} (CP) and Born-Oppenheimer (BO) molecular dynamics (MD), collectively known as {\it ab initio} (AI) MD. In AIMD, the forces are calculated using a first-principles electronic structure method, typically based on density functional theory (DFT). BOMD, used in the present study, minimizes the Kohn-Sham energy functional at each time step, keeping the nuclear positions frozen. In nearly all force field and AIMD simulations of water at ambient conditions there seems to be a strong driving force to form highly directional H-bonds, leading to tetrahedral structures that are in general over-structured in terms of the derived PCFs.  One exception is the coarse-grained mW water model~\cite{molinero2008water}, which has two terms in the interaction potential corresponding to anisotropic tetrahedral interactions and isotropic vdW interactions, respectively, and which gives a maximum peak height of 2.3 in the O-O PCF at room temperature, in close agreement with recent analyses of experimental diffraction results.~\cite{Soper2007, leetmaa2008diffraction, Wikfeldt2009,Fu2009, NeufeindMolPhys,Huang2011} This model was shown to feature fluctuations between tetrahedral and disordered species resulting in a liquid-liquid transition in the supercooled region.~\cite{moore2009growing} Empirical force-field models which have over-structured PCFs in agreement with older determinations~\cite{Soper2000,Hura2000} have however also been shown to exhibit liquid-liquid phase transitions in the supercooled regime, {\it e.g.}, Refs.~\citenum{Poole1992,paschek2008tip4pew,Abascal2010} and~\citenum{brovchenko2005liquid}, indicating that the PCFs are not decisive for general trends in the thermodynamic behavior in water simulations. 

Until recently, AIMD simulations of water have almost exclusively been performed with the BLYP~\cite{BLYP} and PBE~\cite{PBE} exchange-correlation (XC) functionals. However, these functionals are shown to significantly over-structure liquid water,~\cite{Grossman2004} as seen from the maximum value and sharpness of the first peak in the oxygen-oxygen PCF compared to recent data and analyses.~\cite{Soper2007,leetmaa2008diffraction,Fu2009,Wikfeldt2009, NeufeindMolPhys,Huang2011} AIMD simulations of water have furthermore been shown to depend on which functional is applied and to give different predictions for different XC functionals.~\cite{VandeVondele2005} MD simulations performed using functionals based on the generalized gradient approximation (GGA) tend to over-structure liquid water and lead to diffusion constants two to three times too small compared to experiment; using hybrid functionals only marginally improves the results.~\cite{Todorova2006} In addition, it has been shown that PBE-based AIMD simulations lead to a melting point of ice at 417 K and therefore simulations at ambient conditions with this functional will describe a deeply supercooled state which is strongly over-structured with respect to real liquid water at ambient conditions.~\cite{yoo2009dftphase}  As we will show in the following, inclusion of the more isotropic vdW interaction balances the directional forces allowing a partial break-down of the H-bond network and a much less structured liquid. 

\section{Methods}

\subsection{Role of van der Waals (vdW) forces}

Small water clusters have been studied using the PBE and BLYP XC functionals, which do not explicitly include vdW interactions, and the results compared to high accuracy methods such as coupled cluster (CCSD(T)) and M{\o}ller-Plesset (MP2). With PBE~\cite{Ireta2004} near chemical accuracy for the strength of the H-bond for the water dimer is obtained while BLYP consistently underbinds small water clusters.~\cite{Santra2008} However, discrepancies arise and increase with the size of the water cluster for both PBE and BLYP. This has been ascribed to the lack of a description of vdW forces.~\cite{Santra2008} One could thus argue that obtaining the correct result for the water dimer is essential but no guarantee for a correct description since not all physical interactions relevant for larger clusters are sampled by the dimer. 

While it is well established that at low temperatures H-bonds give the major contributing factor to the dynamics and structure of water, vdW interactions have also been suggested to be important~\cite{Cho1997,Schmid2001}. In line with this, thermodynamic considerations have led to the suggestion that at higher temperatures the averaged water interaction potential should resemble that of liquid Argon.~\cite{Lishchuk2010} The angular dependence of the H-bond is anticipated to have a big impact on the PCF and self-diffusion coefficient.~\cite{Michaelides} If, for example, it is too difficult to bend a DFT H-bond, the diffusion coefficient should come out too small, which it does. Many other suggestions to explain the too small diffusion coefficient exist however~\cite{Michaelides} but balancing the directional H-bond interactions with more isotropic vdW forces would intuitively contribute to softening the H-bond network and allow more efficient diffusion. Traditional local and semi-local DFT do not, however, contain non-local vdW interactions, {\it e.g.}, BLYP being especially incapable of describing dispersion.~\cite{Kristyan1994} Influences of vdW interactions have been investigated using MD based on empirical potentials,~\cite{Cho1997,LyndenBell2005} {\it e.g.}, performed with a dispersion-corrected BLYP XC functional,~\cite{Lin2009} or using empirically damped C$_{6}$R$^{-6}$ corrections~\cite{Gianturco1999,Grimme2004,Williams2006,Schmidt2009} to describe the vdW interactions.

A way to introduce vdW forces in DFT from first principles is provided by the van der Waals density functional vdW-DF,~\cite{DFT-vdW} recently used for the first time in AIMD on liquid water.~\cite{Wang2010} The inclusion of vdW forces using the vdW-DF was shown to greatly improve water's equilibrium density and diffusivity. However the vdW-DF MD also produces a collapsed second coordination shell giving rise to new structural problems that have been suggested to depend partially on the choice of exchange used in the vdW functional.~\cite{Wang2010}

The vdW-DF method proposed by Dion {\it et al.}~\cite{DFT-vdW} accounts for exchange by a functional that gives Hartree-Fock-like repulsion at relevant separations and that includes non-local correlation, and thus vdW forces, by calculating the dielectric response in a plasmon-pole approximation. It gives the correct stability trend for low-lying water hexamers~\cite{Kelkkanen2009} but returns a significant underbinding for most H-bonds.~\cite{Gulans2009,Kelkkanen2009,Langreth2009} The underbinding can be remedied by using an exchange functional that gives more binding~\cite{Puzder2006} at typical H-bond separations,~\cite{Gulans2009,Murray2009,Kelkkanen2009} like the PW86,~\cite{PW86} optPBE,~\cite{Klimes2010} and C09~\cite{Co09} exchange functionals. Recently Klimes {\it et al.}~\cite{Klimes2010} proposed a new vdW density functional, optPBE-vdW, based on the original vdW-DF functional.~\cite{DFT-vdW} This scheme shows promise in the description of dispersion and H-bonded systems, as it reduces the underbinding given by the vdW-DF down to chemical accuracy while preserving the correct hexamer trends. However, this improved behavior is obtained at the cost of poorer performance on the binding energy of small molecules.~\cite{Wellendorff} Very recently a  second version of the vdW-DF, called vdW-DF2,\index{} was suggested,~\cite{Lee2010} using a new non-local correlation functional along with a slightly refitted version of the PW86, called PW86R~\cite{Murray2009} as an appropriate exchange functional. Both optPBE-vdW and vdW-DF2 give chemical accuracy for the water dimer, albeit with slightly different balance between non-local-correlation and exchange contributions.  In the present study we therefore wish to investigate the microscopic structure of liquid water by performing AIMD using both the new optPBE-vdW and vdW-DF2 XC functionals to also investigate the importance of the balance between correlation and exchange in liquid water AIMD simulations.

\subsection{The optPBE-vdW and vdW-DF2 exchange-correlation functionals}

In general a vdW-DF functional takes the form
\begin{equation}
E_{xc}=E^{\mathrm{GGA}}_{x}+E^{\mathrm{LDA}}_{c}+E^{\mathrm{nl}}_{c},
\end{equation}
where $E^{\mathrm{GGA}}_{x}$ is an exchange functional using the generalized gradient approximation (GGA), $E^{\mathrm{LDA}}_{c}$ accounts for the local correlation energy by using the local density approximation (LDA). LDA is chosen to avoid double counting of correlation. The non-local correlation energy describing the vdW interaction is given by the six-dimensional integral~\cite{DFT-vdW}
\begin{equation}
E_{\mathrm{c}}^{\mathrm{nl}}=\frac{1}{2}\int \int n({\bf r})\phi ({\bf r},{\bf r}')
n({\bf r}') \mathrm{d}{\bf r}\mathrm{d}{\bf r}',
\end{equation}
where $\phi ({\bf r},{\bf r}')$ is the interaction kernel and depends on the density and its gradient. The non-local term is calculated as suggested in Ref. \citenum{RomanPerez2009}. In the original vdW-DF from Dion {\it et al.} the exchange functional from revPBE~\cite{Zhang1998} is utilized.

The optPBE-vdW functional is constructed like vdW-DF~\cite{DFT-vdW} but uses an alternative exchange functional. The latter takes the same form as both the PBE and RPBE exchange, but the parameters of the exchange enhancement factor are optimized against the S22 database.~\cite{Klimes2010} The S22 database~\cite{Jurecka2006} is a set of 22 weakly interacting dimers, mostly of biological importance, including the water dimer.

The vdW-DF2~\cite{Lee2010} has the form of Eq. (1) and uses the PW86 exchange,~\cite{Wang1986} which is argued in Ref.~\citenum{Murray2009} to give the most consistent agreement with Hartree-Fock (HF) exact exchange, and with no spurious exchange binding. Furthermore, a new approximation for E$_{c}^{\mathrm{nl}}$ is used to calculate the value of the interaction kernel in Eq. (2).~\cite{Lee2010} This new functional has been shown to give very accurate results for the water dimer as compared to benchmark CCSD(T) calculations~\cite{Molnar2009,Lee2010} and to compare closely to the S22 benchmark.~\cite{Takatani2010}

\subsection{Computational protocol}

{\it Ab initio} molecular dynamics simulations are performed in the NVE ensemble with optPBE-vdW, vdW-DF2, and PBE, using the grid-based real-space projector augmented wave GPAW code.~\cite{GPAW2010,GPAW2005} A wave function grid spacing of 0.18 {\AA} and Fermi smearing with a width of $0.01$ eV have been used. The grid spacing has been determined by comparing DFT calculations of water hexamers with CCSD(T) results. In the electronic structure calculations a strict energy convergence criterion of $10^{-7}$ eV per electron is used in order to determine the forces adequately.

All internal bond lengths are kept fixed at 0.9572 {\AA} (an MP2 optimized gas phase geometry obtained from the G2-database)~\cite{G2} but angles are allowed to vary ({\it i.e.} bending vibrations are included); eliminating the high-frequency OH-stretch allows longer time steps in the simulations albeit introducing some uncertainty~\cite{Leung-PCCP2006} which, however, is not relevant for the large differences observed in our simulations between the PBE on the one hand and the vdW functionals on the other since all simulations have this constraint imposed.  In the initial configuration 64 water molecules are placed in a simple cubic lattice with random orientations in a cubic periodic box with side lengths 12.42 {\AA}, to reproduce a water density of 1 g/cm$^{3}$. The geometry is then optimized to obtain a configuration at zero Kelvin (using PBE), from which the MD is started giving the atoms random velocities according to a Maxwell-Boltzmann velocity distribution corresponding to two times 300K, keeping the center of mass of the box stationary. Approximately half of the kinetic energy converts to potential energy thus giving an average temperature around 300K. An initial equilibration of 10 ps using the PBE XC functional is performed followed by 2.5 ps vdW equilibration of the simulations using optPBE-vdW and vdW-DF2. For all methods equilibration was followed by production runs for 10 ps which is the minimum time reported necessary due to the slow diffusion of water.~\cite{Sorenson2000} Using 64 water molecules has been shown to be adequate to remove the most significant problems concerning finite size effects~\cite{Head-Gordon2002} and is feasible within the current computational capabilities. The Verlet algorithm is employed using a time step of 2 fs in the NVE ensemble. Using this type of ensemble the temperature is allowed to fluctuate and the average temperature of the PBE, vdW-DF2 and optPBE-vdW simulations were 299K, 283K and 276K, respectively. The same computational setup has been used for the PBE and vdW density functional MD simulations in order to allow direct comparison of the different models. Since simulations with PBE at ambient conditions describe a deeply supercooled state relative to its melting point at 417 K~\cite{yoo2009dftphase} the PBE simulations are only performed here to provide a reference for the effects of including vdW interactions through the optPBE-vdW and vdW-DF2 functionals.

\begin{figure}[h]
  \centerline{\hbox{ \hspace{0in}
\includegraphics[angle=0,scale=0.22]{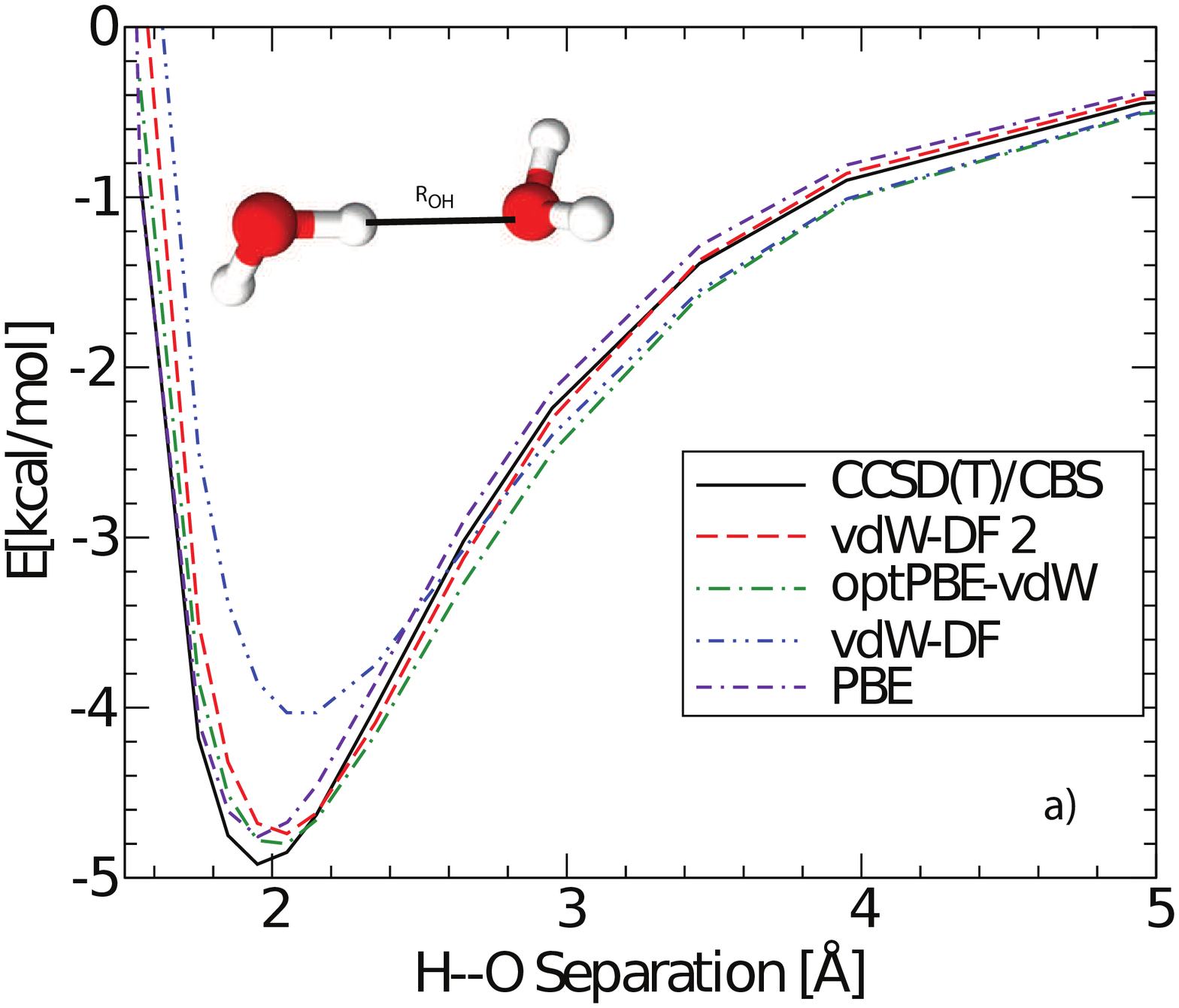}
\includegraphics[angle=0,scale=0.20]{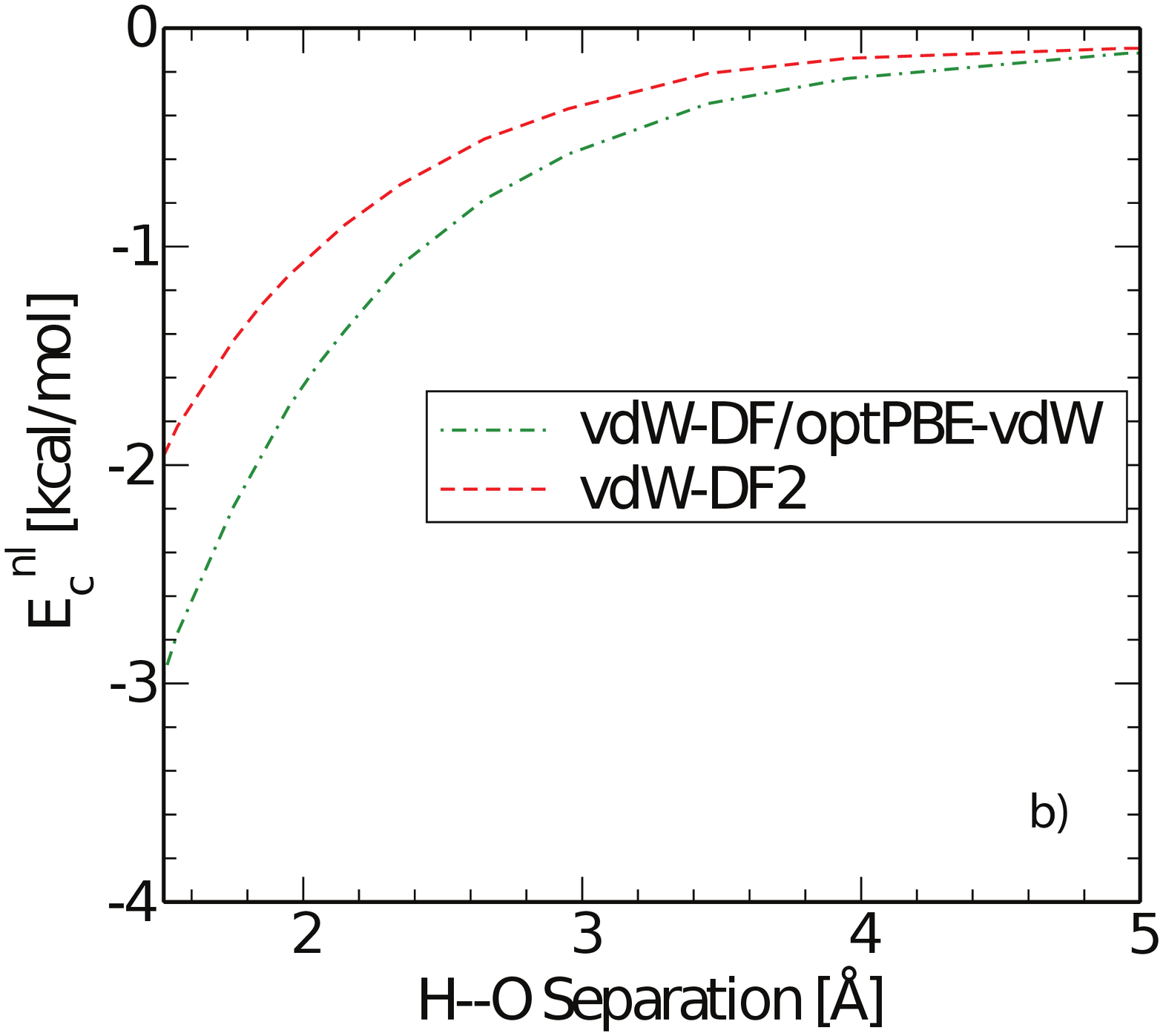}}}
\caption{a) The water dimer potential energy curves calculated with DFT using the XC functionals PBE, vdW-DF, vdW-DF2 and optPBE-vdW, respectively, are compared to CCSD(T)/CBS wave function results.~\cite{Molnar2009} b) The distance dependence of the non-local contribution (Eq. (2)) to the interaction energy of the water dimer for the XC functionals vdW-DF2 and optPBE-vdW, shows that, when they give similar potential energies (Fig. 1a)) they do so for different reasons; the optPBE-vdW gets more binding from a stronger vdW attraction but vdW-DF2 gets more net attraction from a less repulsive exchange.}
\label{fig:dimer}
\end{figure}

\section{Results}

\subsection{Water dimer}

Before discussing the MD results we compare the functionals for a simpler but still relevant system: the water dimer. Fig. \ref{fig:dimer} a) illustrates the potential energy curve for the water dimer calculated using PBE, vdW-DF, vdW-DF2 and optPBE-vdW in comparison with the benchmark CCSD(T) curve from Ref. \citenum{Molnar2009}. Fig. \ref{fig:dimer} a) shows that the vdW functionals are capable of describing this basic constituent of liquid water extremely accurately, however for different reasons. The non-local contribution (E$_{c}^{\mathrm{nl}}$) to the dimer binding from the two functionals is plotted in Fig. \ref{fig:dimer} b).  The non-local part of the optPBE-vdW functional, which is based on the older approximation, is more attractive as mentioned in Ref. \citenum{Lee2010}. Since less attraction stems from the non-local interaction in the vdW-DF2, while the total energy for the dimer is almost identical to that of optPBE-vdW, the remaining part of the interaction energy must give a larger contribution for the vdW-DF2 than for optPBE-vdW. The remaining part of the interaction energy includes electrostatic interaction, electronic correlation, and more or less repulsive exchange. Since electrostatic interactions only depend on separation, and local correlation is treated identically with the LDA correlation in both cases, this difference has to come from the different choices for the exchange. The PW86 exchange in vdW-DF2 is hence less repulsive than the optPBE exchange in optPBE-vdW; a possible cause of the reported collapsed second-shell structure was in Ref. \citenum{Wang2010} suggested to be that the non-local parameterization of exchange used in vdW-DF and optPBE-vdW may be too attractive when used in MD. This is, however, not the case, as seen from the pair-correlation functions (PCFs), to be discussed next.

\subsection{Pair-correlation functions}

\begin{figure}[h]
  \centerline{\hbox{ \hspace{0in}
\includegraphics[angle=0,scale=0.23]{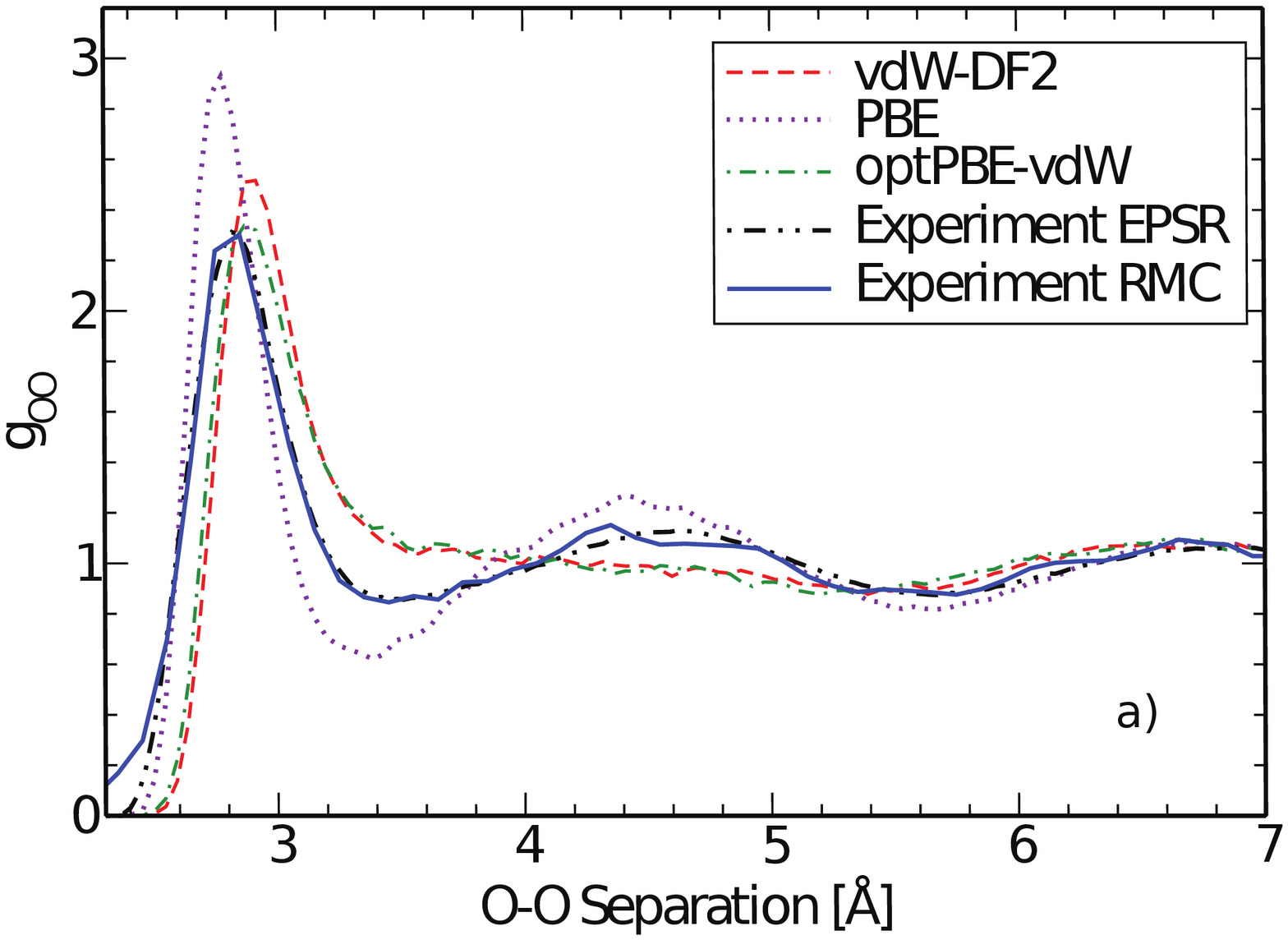}
\includegraphics[angle=0,scale=0.23]{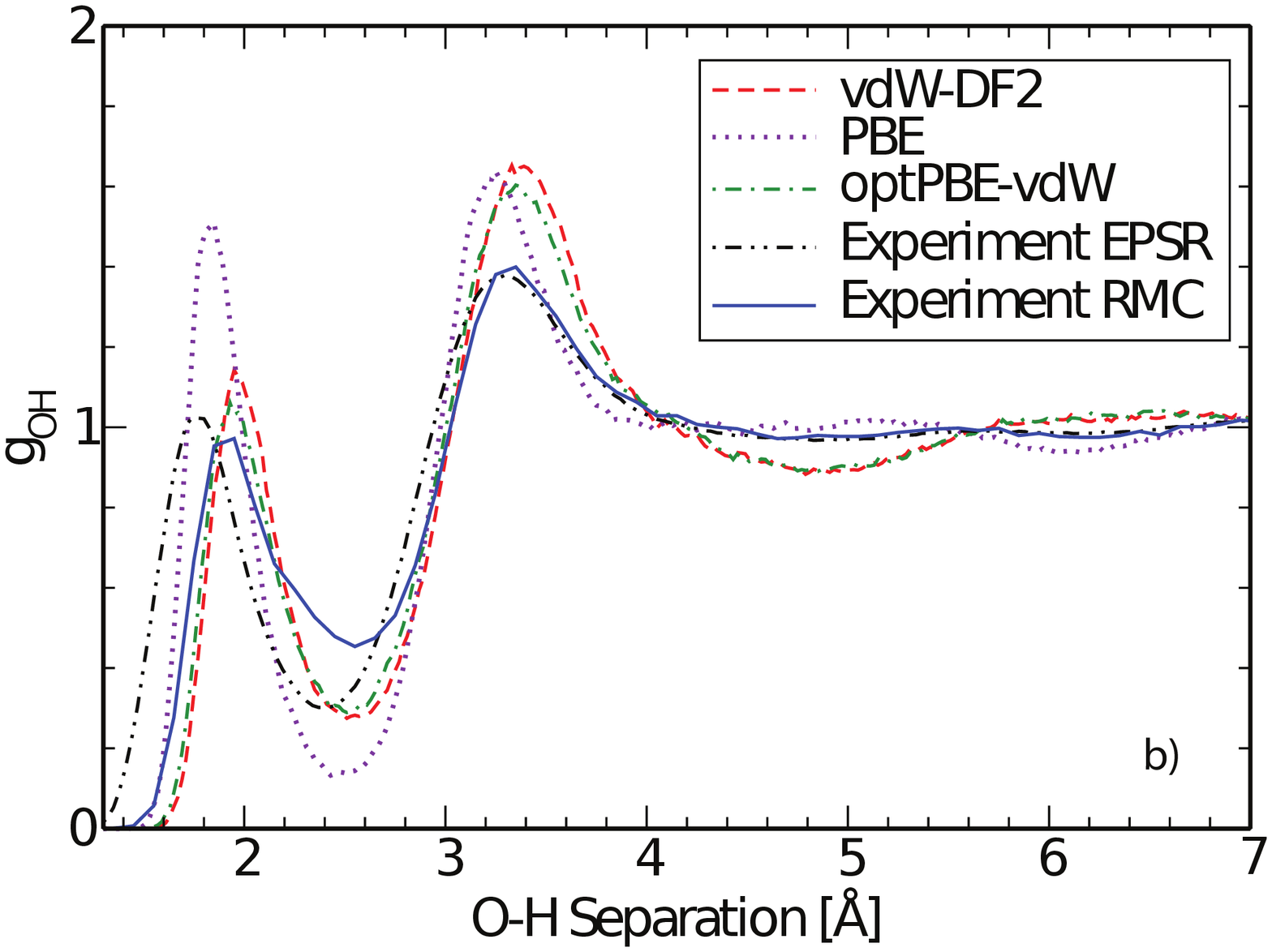}}}
\caption{a) Oxygen-oxygen PCFs (g$_{\mathrm{OO}}$) obtained from experimental data using EPSR~\cite{Soper2007} and RMC~\cite{Wikfeldt2009} in comparison with PCFs obtained by DFT MD simulations using PBE, optPBE-vdW and vdW-DF2. b) Oxygen-hydrogen  PCFs (g$_{\mathrm{OH}}$) obtained from experimental data using EPSR~\cite{Soper2007} and RMC~\cite{Wikfeldt2009} in comparison with PCFs from PBE, optPBE-vdW and vdW-DF2.}
\label{fig:RDF_OH}
\end{figure}

Fig. \ref{fig:RDF_OH} a) illustrates that AIMD simulations of liquid water using vdW-DF2 and optPBE-vdW give very similar O-O PCFs which are, however, very different from the O-O PCF from PBE and furthermore from those derived from experiment using either Empirical Potential Structure Refinement (EPSR)~\cite{Soper2007} or Reverse Monte Carlo (RMC)~\cite{RMC} to fit the structure factor.~\cite{leetmaa2008diffraction,Wikfeldt2009} In the simulations both functionals result in the same characteristics as reported in Ref. \citenum{Wang2010}, including a lower first peak shifted to larger O-O separation than for normal GGAs as well as for experiment on ambient water. The second coordination shell at 4.5 {\AA} is also completely smeared out where correlations from the region 4-5 {\AA} have instead moved into the region 3.3-3.7 {\AA}. The non-local correlation differences in the functionals do not, however, result in significantly different O-O PCFs, but we note a somewhat higher (2.5) first peak for vdW-DF2 compared to optPBE-vdW (2.3). Since the latter gives a slightly stronger non-local contribution we take this as indication that it is indeed the vdW contribution that so strongly affects the first shell structure in the simulations. 

In contrast, the very recent vdW-DF MD simulation showed that by changing the exchange in vdW-DF from revPBE to PBE, the second shell structure again became well defined.~\cite{Wang2010} However, the exchange functionals of revPBE and PBE are quite different, making an explanation in terms of the exchange less likely; the potential energy curve of the dimer is furthermore not reproduced very well using the PBE exchange with LDA and non-local correlation suggesting that substituting revPBE by PBE for the exchange does not lead to consistent improvement in the description.

Compared to the experimentally derived O-O PCFs it is clear that the PCF obtained from PBE is severely over-structured while the simulations including vdW forces have resulted in a significantly less structured PCF than what is experimentally observed for ambient liquid water. Clearly neither simulation model gives direct agreement with the experimental O-O PCF even though, in the case of the vdW functionals, small water clusters are described very accurately. We will address this aspect in the discussion section below.

Some discrepancy in the O-H PCF for the vdW XC functionals compared to experiment is seen in Fig. \ref{fig:RDF_OH} b), which we shall now consider. The O-O correlations can be obtained from a Fourier transform of the x-ray diffraction data, if a large enough k-range has been measured and the data can be properly normalized; x-ray scattering is strongly dominated by the electron-rich oxygens.  Neutron diffraction data, on the other hand, contain simultaneous information on the H-H, O-H and to some extent, the O-O PCFs, making a direct Fourier transform to extract a specific PCF inapplicable. Various fitting schemes of structure models to the experimental structure factors have therefore been developed, and we show two such fits to the same experimental data using the EPSR~\cite{Soper2007} and RMC method,~\cite{Wikfeldt2009} respectively. 

There is a significant difference in the first peak position in the O-H PCF between the EPSR and RMC fits compared to what is found for the O-O PCF. This can be understood from the relatively lower sensitivity of the neutron data to specifically the O-H correlation in comparison to the sensitivity of x-ray data to the O-O correlation.~\cite{Wikfeldt2009} The EPSR technique uses the assumed reference pair-potential to provide structural aspects not included in the experimental data,~\cite{Soper2005Asym,Wikfeldt2009} while structural aspects not determined by the experimental data, or imposed constraints, will in the RMC technique simply result in a phase-space weighted sampling of structures consistent with the experimental structure factors;~\cite{Jedlovszky} combining the two methods thus gives additional information on the uncertainties and assumptions in the resulting fits. It is interesting to note that the RMC method gives a shift in the first peak of the O-H correlation out to nearly 2 {\AA},~\cite{Wikfeldt2009} which agrees well with the vdW MD simulations presented here, while the EPSR solution is closer in position to the PBE, likely reflecting the SPC/E starting force-field in the EPSR fitting procedure. Note, that both the RMC and EPSR fits reproduce the experimental scattering data equally well, implying that the position of the first intermolecular OH correlation is not strictly determined by the data, which leaves an uncertainty in the diffraction-derived O-H PCF.~\cite{pusztai1999partial, Wikfeldt2009}  The first peak in the PBE O-H PCF is clearly too high and the first minimum at 2.5 {\AA} too low, however, while all three simulations exaggerate the height of the second peak at 3.2~-~3.6 {\AA}; this can, however, be expected to be reduced by including quantum effects, {\it e.g.}, Ref.~\citenum{Paesani2007}.

\subsection{Angular distribution functions and hydrogen-bonding analysis}

The van der Waals functionals provide a smoother angular structure with less tetrahedral bonding as demonstrated by the angular distribution functions and the average number of H-bonds per water molecule; here we use the cone criterion from Ref. \citenum{Wernet2004} as a geometric H-bond definition:
$r_{\mathrm{OO}} < r_{\mathrm{OO}}^{\mathrm{max}} - 0.00044 \delta_{\mathrm{HOO}}^2$. This defines a cone around each H-bond-donating OH group, where $r_{\mathrm{OO}}^{\mathrm{max}}=3.3$ {\AA} is the maximum OO distance at zero angle $\delta_{\mathrm{HOO}}$, where $\delta_{\mathrm{HOO}}$ is the H-O$\cdots$O angle quantifying the angular distortion of the H-bond.
Table \ref{HB} shows the H-bond statistics for PBE, optPBE-vdW and vdW-DF2.
PBE is seen to prefer a tetrahedral H-bond coordination with a majority of the molecules having 4 H-bonds. Including non-local correlation has a large effect where, for both optPBE-vdW and vdW-DF2, the H-bond distribution shifts from a majority with four H-bonds to instead a predominance of species with two or three. Comparing the two vdW functionals we observe that the optPBE-vdW has a slightly larger amount of water molecules having two or three H-bonds compared to vdW-DF2 which we ascribe to the relatively more repulsive exchange and stronger non-local contribution in the former. The vdW-DF2 with its relatively weaker vdW interaction shows slightly higher preference to forming H-bonds. This analysis suggests that there is a competition between isotropic vdW forces and directional H-bonds, resulting in fewer or more H-bonds per water molecule depending on the applied approximations; however, between the vdW models the average number of H-bonds varies only weakly despite differences in vdW strength.

\begin{figure}[h]
\includegraphics[angle=0,scale=0.38]{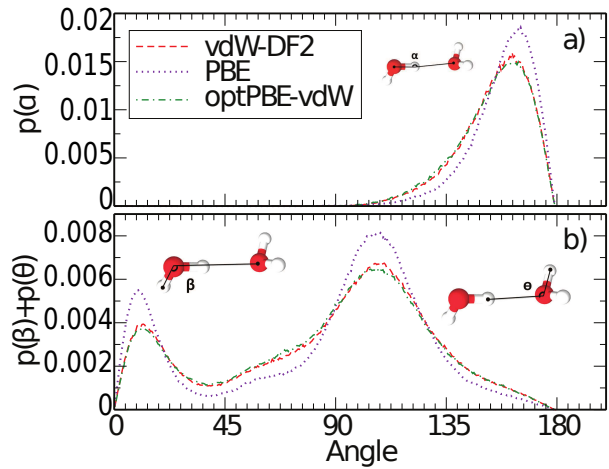}
\caption{Molecular angular distributions in liquid water according to MD simulations using DFT with the indicated functional. a) The angular distribution functions of the O-H$\cdots$O angle. b) the H-O$\cdots$O (first peak) and $\theta=<$H$\cdots$O-H (second peak) angles obtained using the
XC functionals PBE (yellow), vdW-DF2 (red) and optPBE-vdW (green). When including vdW interactions a softening of the structure is seen from the broader distribution of
angles.}
\label{fig:ADF}
\end{figure}

\begin{table}[h]
\begin{tabular}{|c|c|c|c|}\hline
No. H-bonds $\backslash$ Method & PBE & optPBE-vdW & vdW-DF2 \\
\hline\hline
1  & 2 &    10 &    8   \\ \hline
2  & 12 &   29 &    27  \\ \hline
3  & 31 &   37 &    38  \\ \hline
4  & 52 &   22 &    25  \\ \hline
5  & 3 &    1  &    1   \\
\hline
\end{tabular}
\caption{Percentage distribution of hydrogen bonds per water molecule calculated using the cone criterion from Ref. \citenum{Wernet2004}. PBE results are shown to favor more H-bonds compared to either vdW functional, which both allow for a larger number of molecules to break the tetrahedral structure with four bonds.}
\label{HB}
\end{table}

We note in particular the low number of double-donor, double-acceptor tetrahedral molecules according to the cone criterion\cite{Wernet2004} for the two vdW models. In fact, the large number of broken H-bonds in the vdW simulations suggests that these models are in closer agreement with predictions from x-ray spectroscopies~\cite{Myneni2002,Wernet2004, tokushima2008high, PNAS2009,Nilsson2010,Tokushima2010,Leetmaa2010} compared to most other AIMD models and future calculated x-ray spectra based on optPBE-vdW and vdW-DF2 structures may provide an interesting opportunity to obtain further insight regarding the interpretation of these spectra.

The angular distribution functions (ADFs) of the H-bonds are shown in Fig. \ref{fig:ADF}. The ADFs of H-bond acceptor and donor give information on the orientational flexibility of the water molecules. In the ADFs only the angles between a central molecule and the molecules of the first solvation shell are considered by using a cutoff distance corresponding to the first minimum in the PBE O-O PCF; this distance was applied also to the vdW MDs where the second shell is smeared out and no minimum is visible. Fig. \ref{fig:ADF} a) displays the distributions of donor angles $\alpha=<$O-H$\cdots$O for the various simulations. The first peak in Fig. \ref{fig:ADF} b) is $\beta$, the deviation of the O-H$\cdots$O bond from being linear, which gives information on the flexibility of donor H-bonds, while the second peak is the acceptor angle $\theta=<$H$\cdots$O-H. The distribution of angles has been found to depend on the choice of water model.~\cite{Lee2006} The picture of a competition between non-directional vdW interactions and directed H-bonds seems to be supported by the ADFs as illustrated by the fact that the model without vdW forces (PBE) has no incentive to deviate from a structure of strong H-bonds, thus resulting in a relatively straight H-bond angle. When including vdW forces the H-bonds become significantly more bent. In general a softening of the structure is seen from the broader ADFs obtained in case of the vdW-DFs.

\subsection{Tetrahedrality and asphericity}
Two useful measures of the local coordination of molecules in water are the tetrahedrality~\cite{chau1998tetra,errington2001tetra} and asphericity~\cite{ruocco1992voronoi} parameters. The former quantifies the degree of tetrahedrality in the nearest neighbor O-O-O angles and is defined as
\begin{equation}
Q = 1 - \frac{3}{8}\sum_{i=1}^{3}\sum_{j>i}^{4}\left( \cos \theta_{i0j} + \frac{1}{3} \right) ^2
\label{eq:tetra}
\end{equation}
where $\theta_{i0j}$ is the angle formed by two neighboring oxygen atoms $i$ and $j$ and the central molecule $0$. Only the four nearest neighbors are taken into account which makes $Q$ a very local measure. Perfect hexagonal ice gives $Q=1$ for all molecules while the ensemble average over an ideal gas gives $<Q>=0$.~\cite{errington2001tetra}
The asphericity parameter is defined as 
\begin{equation}
\eta = \frac{A^3}{36\pi V^2} 
\label{eq:eta}
\end{equation}
where A and V are the area and volume of the Voronoi polyhedron of the molecule in question. Contrary to $Q$, $\eta$ is sensitive also to interstitial molecules outside the first shell and to the second coordination shell since these add surfaces to the Voronoi polyhedron, making it more spherical. The two relevant limits for water are that of hexagonal ice, which gives $\eta=2.25$, and that of a perfect sphere which gives $\eta=1$; larger disorder in the local coordination thus gives smaller values of $\eta$.

\begin{figure}[h]
  \centerline{\hbox{ \hspace{0in}
\includegraphics[angle=0,scale=0.25]{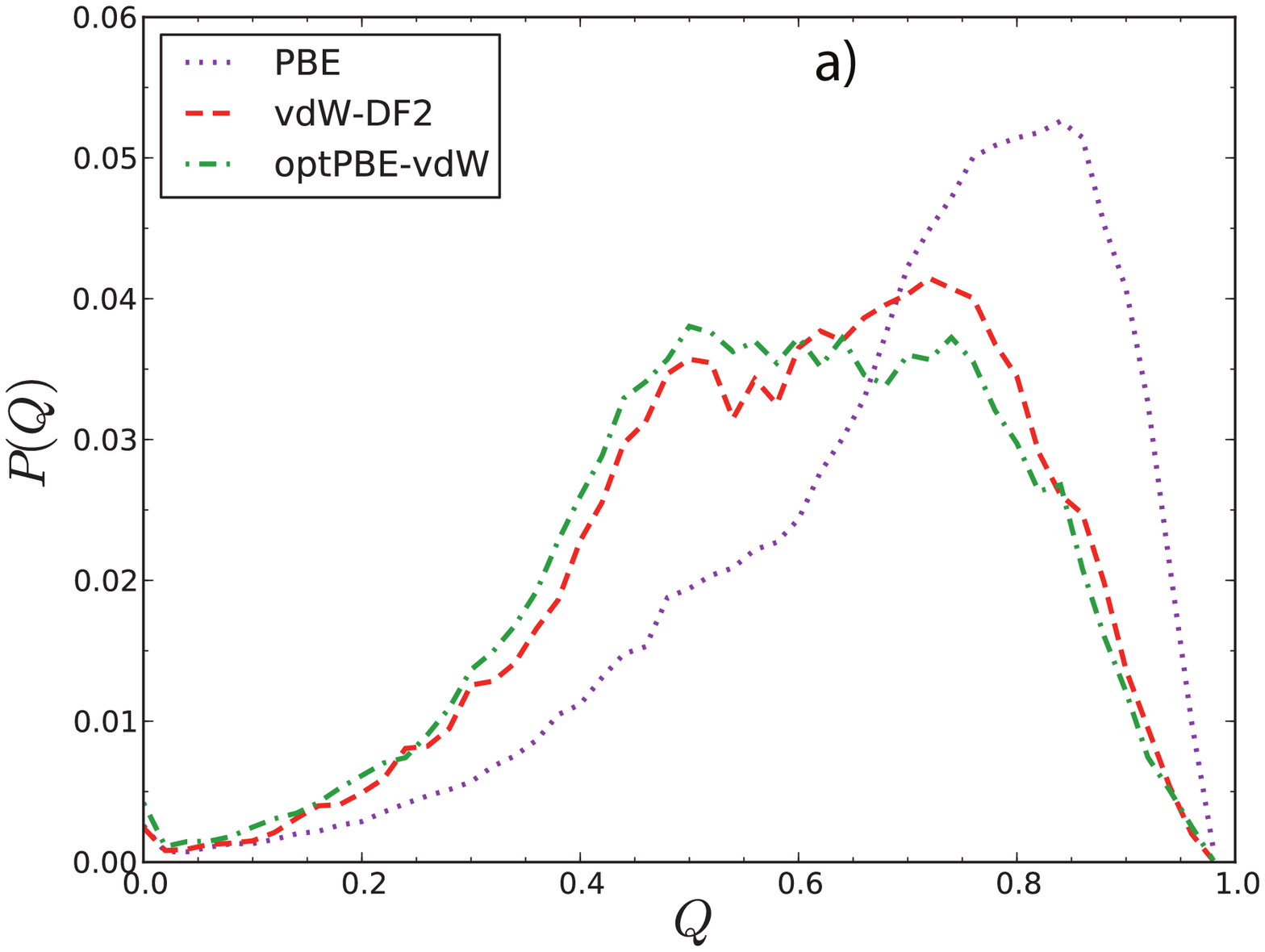}
\includegraphics[angle=0,scale=0.25]{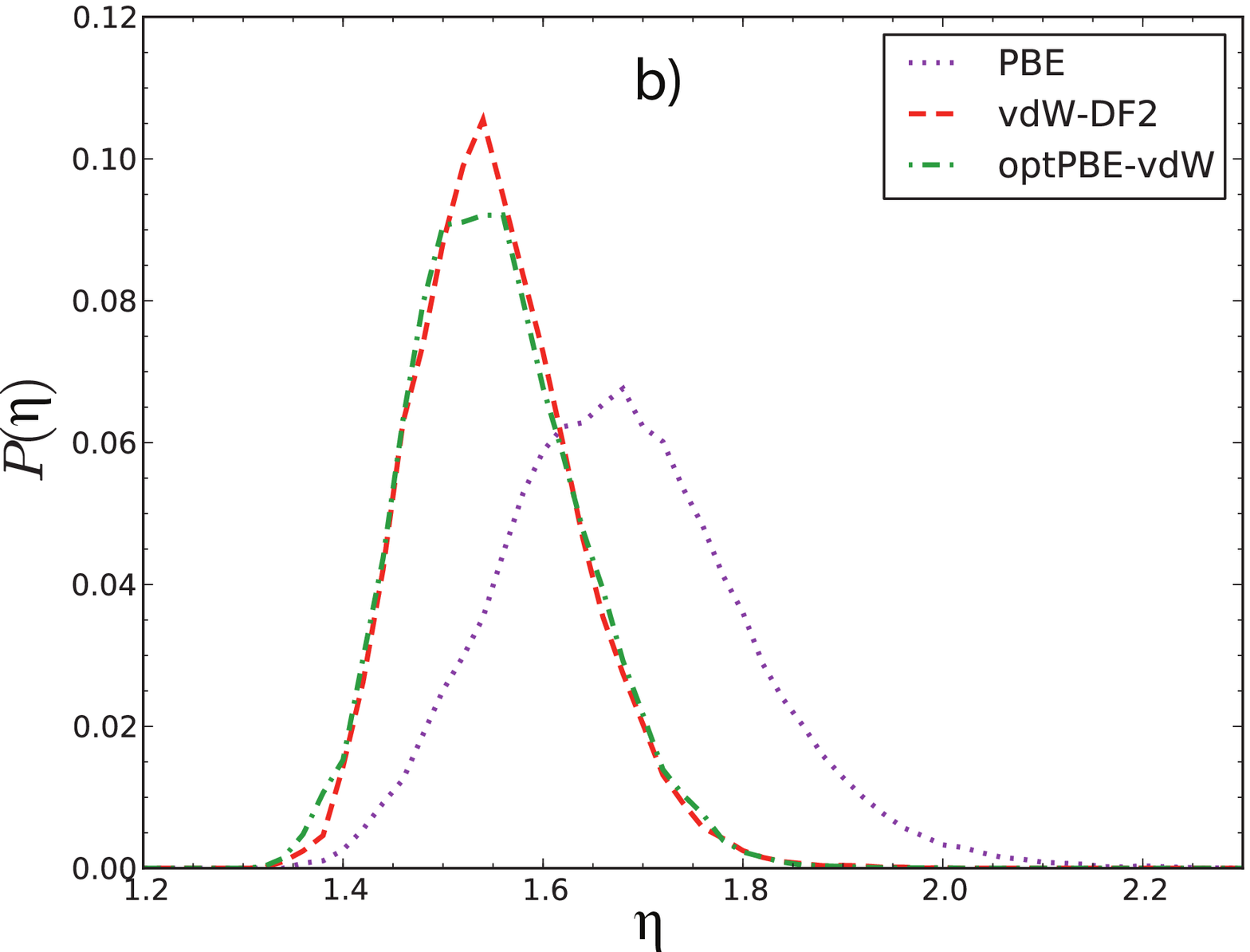}}}
\caption{a) Distributions of the tetrahedrality parameter $Q$. vdW interactions lead to significantly lower average tetrahedrality and a strong low-$Q$ peak from interstitial molecules around $Q=0.5$. b) Distributions of the asphericity parameter $\eta$. A large effect of vdW interactions is seen with a shift towards more spherical (less ice-like) values.}
\label{fig:tetra_eta}
\end{figure}

As Fig. \ref{fig:tetra_eta} shows, the inclusion of the vdW interaction not surprisingly has a dramatic effect on both the tetrahedrality and asphericity distributions. The PBE simulation displays a strong peak at $Q=0.8$, signifying a dominance of locally tetrahedral O-O-O angles, while both vdW simulations show an attenuation and shift of the high-$Q$ peak to lower tetrahedrality along with the appearance of a strong low-$Q$ peak associated with interstitial molecules at non-tetrahedral positions between the first and second coordination shells. Out of the two vdW models, optPBE-vdW is seen to be somewhat less tetrahedral, consistent with their differences in H-bond statistics and PCFs discussed above.  This is clearly illustrated by the average tetrahedrality which is 0.692, 0.602 and 0.583 for PBE, vdW-DF2 and optPBE-vdW, respectively. In comparison the average tetrahedrality has been estimated to be 0.576 using the EPSR method;~\cite{Soper2008} note, however, that the tetrahedrality parameter is experimentally rather uncertain, {\it i.e.} the same diffraction data have been shown to support tetrahedrality values ranging from 0.488 to 0.603.~\cite{Wikfeldt2009} 

An even larger difference is seen in the asphericity distributions; the two vdW models show sharper peaks centered at lower asphericity values compared to PBE. This directly reveals the large disorder in second-shell correlations in the vdW models, resulting from the tendency to form more isotropic local structures when vdW forces are included. Similarly to the comparison between the PCFs of the two vdW models discussed above, it can be seen here that despite non-local differences between vdW-DF2 and optPBE-vdW their respective liquid water structures turn out to be rather comparable in terms of both first- and second-shell correlations. The average asphericity is 1.681, 1.552 and 1.552 for PBE, vdW-DF2 and optPBE-vdW, respectively. 

\section{Discussion}

Considering the accuracy of the present versions of non-local correlation functionals, as calibrated against benchmark CCSD(T) and MP2 calculations for water dimer, water hexamers,~\cite{Kelkkanen2009} the S22 database,~\cite{Klimes2010} we will here explore the possibility that the interactions between molecules in the simulation box are given sufficiently accurately by the functionals and that the resulting discrepancy between simulated and observed O-O PCF is rather due to limitations and constraints in the simulation protocol.

Comparison of the results from the simulations using PBE with those including the vdW interactions shows a strong shift in the balance between directional H-bonding and more isotropic interactions; the former leads to tetrahedral H-bond coordination and low density while the latter favors a more close-packed ordering and higher density, as evidenced by the loss of distinction between first and second coordination shells and the reduced number of H-bonds. The simulations have in all cases been performed with internal OH distances fixed to the gas phase value; eliminating the high-frequency OH stretch allows longer time-steps to be used in the AIMD, but not allowing the internal OH distance to vary according to H-bond situation has been shown to lead to somewhat less structured PCFs in earlier work.~\cite{Leung-PCCP2006} However, since the simulations with PBE, optPBE-vdW and vdW-DF2 were all run with the same constraint in terms of internal OH distance this cannot explain the large effects on the O-O PCF from including the vdW non-local correlation.

We note that recent, high-precision x-ray diffraction measurements~\cite{Huang2011} of ambient (25$^{\circ}$C) and hot (66$^{\circ}$C) water resolve shell-structure out to $\sim$12 {\AA} in agreement with conclusions from SAXS~\cite{PNAS2009}; shell structure out to the fifth neighbor distance has been resolved before but only for supercooled water.~\cite{NeufeindMolPhys,Yokoyama2008} Based on analysis of large-scale simulations with the TIP4P/2005 force-field~\cite{Abascal2005}  the shell structure could be assigned as due to an instantaneous LDL-like minority species.~\cite{Huang2011} The observed spatial extent of the correlation (12 {\AA}) is similar to the size of the present simulation box (12.42 {\AA}) making it unlikely that the simulation box is sufficiently extended to support such experimentally observed instantaneous structures. 

\begin{figure}[h]
\includegraphics[angle=0,scale=0.4]{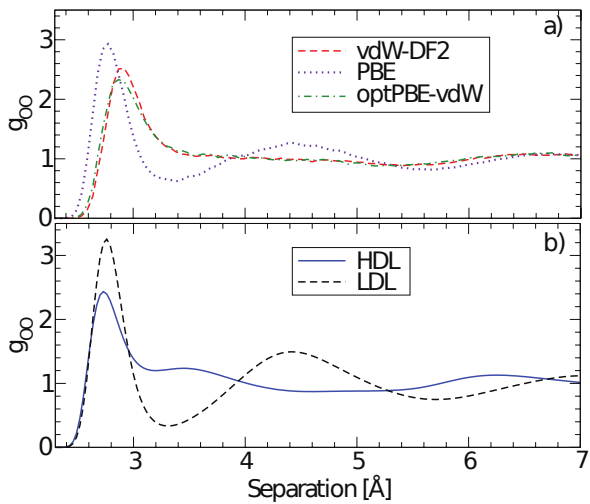}
\caption{a) Oxygen-oxygen PCFs (g$_{\mathrm{OO}}$) obtained by MD simulations from DFT with PBE, optPBE-vdW and vdW-DF2  functionals. b) Experimental PCFs for high- and low-density water.~\cite{Soper2000}}
\label{fig:RDF_OO}
\end{figure}

We compare the PCFs from the simulations performed using the vdW-DF2 and optPBE-vdW XC functionals  (Fig. \ref{fig:RDF_OO} a)), respectively, with the results of a neutron diffraction study~\cite{soper2000highdens} where LDL and HDL O-O PCFs were extrapolated from data at different pressures; the resulting PCFs are shown in Fig. \ref{fig:RDF_OO} b). The EPSR derived HDL PCF is rather similar to the PCF obtained using a Fourier transform of x-ray diffraction data at high pressures~\cite{Okhulkov1994} and furthermore seen to be very similar in terms of the second- and third-shell structure to that derived from vdW-DF2 and optPBE-vdW MD simulations; the effect of increasing pressure on the O-O PCF is that the 4.5 {\AA} correlation disappears and moves to the 3.3~-~3.7 {\AA} region and the third shell is shifted down to 6 {\AA}.~\cite{Okhulkov1994} The O-O PCFs obtained using the vdW functionals similarly show a lack of well defined structure at 4.5 {\AA}, an increase in correlations at 3.3~-~3.7 {\AA} and show a shift towards shorter separations in comparison to PBE of the correlation at 6~-~6.5 {\AA}, as is seen from Fig. \ref{fig:RDF_OO}. Both are clear indications towards HDL water. However, in contrast to the high pressure PCFs, a well defined peak at 3.5 {\AA} is not present in the vdW MD simulations, but only an increase in correlations, and the first peak position is shifted outwards, which is not observed for pressurized water. 

Assuming that the AIMD simulations with non-local correlation and more isotropic interactions have led to a more compact, HDL-like structure, it could be argued that a well-defined peak at 3.5 {\AA} should not be expected since, as deduced from XES spectra at different temperatures~\cite{tokushima2008high,PNAS2009}, HDL-like water at ambient conditions should be thermally excited with a more expanded first shell and therefore further disordered in comparison to HDL water obtained under pressure. In particular, entropy effects due to thermal excitations leading to higher disorder can be expected to create a structure where both the first shell and, in particular, the collapsed second shell are distributed over a range of distances, leading to molecules in what is often denoted interstitial positions and with the first O-O peak appearing at longer distance when not under pressure. In this respect a comparison with the amorphous high-density (HDA) and very high-density (VHDA) ices is of interest, where, for VHDA, the second shell moves inwards and a peak at 3.4 {\AA} develops while for HDA a peak is found at 3.7 {\AA} and the second peak broadens significantly. This indicates that various interstitial sites may be occupied making the high-density forms less well-defined.~\cite{Finney2002,Koza2005,Koza2006,Schober1998,Tulk2002,Venkatesh1974} It should be mentioned that a peak at {$\sim$}3.7 {\AA} is present in the MD simulation performed by Wang {\it et al.}~\cite{Wang2010} using the earlier vdW-DF ~\cite{DFT-vdW} formulation of the functional.

If we consider the proposed model of fluctuations between HDL and LDL,~\cite{PNAS2009,huang2010increasing} it could well be that the vdW models under the present conditions only generated HDL-like structures while without including vdW the resulting structure is clearly more LDL-like. Having two balancing interactions that favor opposite structural properties is a prerequisite for fluctuations; it is clear that by tuning either the importance of H-bonding or the vdW interaction the preference for either structure will be affected in the simulations. However, if the two proposed structures of liquid water truly do coexist as endpoints of fluctuations in nanosized patches of different local density, as suggested in Ref. \citenum{PNAS2009}, then an AIMD with only 64 water molecules in a fixed volume may not be suitable to observe this behavior; a much larger box size and an NPT ensemble simulation allowing the box size to vary would be required. The relatively small simulation (12.42 {\AA} box length) and short run time (10 ps) may only observe a local structure of water which, in this picture, is either approximating LDL- or HDL-like. It should be noted that the simulations are run in the NVE ensemble with density fixed to correspond to ambient conditions which, under the assumption that ambient water is dominated by HDL, should furthermore favor an HDL-like structure over fluctuations towards LDL, if energetically allowed, as seems to be the case with vdW interactions included. 

Exploring the hypothesis that the experimentally measured O-O PCF in reality is the result of a spatial or temporal average over fluctuating structures as suggested in, {\it e.g.}, Ref. \citenum{PNAS2009}, and that the vdW-DF2 and optPBE-vdW functionals actually provide a sufficiently accurate interaction potential, we will consider what additional contribution would be required to achieve agreement with the measured O-O PCF. The PCFs are, however, not directly measurable but derived from experimental data and we first need to discuss specifically the choice of O-O PCF for the comparison since different reference PCFs are used in the literature.

X-ray and neutron diffraction data treated in conjunction, either by the technique of empirical-potential structure refinement (EPSR)~\cite{Soper2007} or by reverse Monte Carlo (RMC) simulations,~\cite{leetmaa2008diffraction,Wikfeldt2009} as well as from directly Fourier transforming the latest high-quality x-ray diffraction data sets~\cite{Fu2009,NeufeindMolPhys,Huang2011} and the early data set of Narten and coworkers~\cite{Narten1971, Narten1970} all give a broad and slightly asymmetric first O-O peak with height 2.1-2.3 which is significantly lower than from standard MD simulations (height $\sim$3) and from previous analyses of either only neutron diffraction data using EPSR~\cite{Soper2000} or analysis of the total x-ray scattering I(k) in terms of comparison to computed I(k) from MD simulations.~\cite{Hura2000,Hura2003}

There were, however, problems with both the latter approaches~\cite{Soper2000, Hura2000,Hura2003} since neutron diffraction mainly measures H-H and O-H correlations and thus contain insufficient information to modify the initial SPC/E force-field guess in EPSR to a solution that also describes the O-O PCF, which is mainly determined by x-ray diffraction. The assumption by Hura {\it et al.}~\cite{Hura2000,Hura2003} was that some existing MD force-field should describe the total I(k); the best agreement was found for the TIP4P-pol2 potential from which pair-correlation functions were subsequently extracted to represent experiment.  However, the internal molecular scattering strongly dominates I(k) in x-ray scattering and masks the more relevant intermolecular scattering such that small, but significant discrepancies in phase and amplitude at higher k,~\cite{leetmaa2008diffraction} which determine the shape and height of the first O-O peak, were not observed and taken into account. Since the two independent studies based on, respectively, neutron and x-ray diffraction data arrived simultaneously at similar peak height and shape this was understandably taken as proof that the O-O PCF had been determined correctly; however, both studies reproduced in a sense the force-field used for the analysis and neither was strictly correct. 

This state of affairs was analyzed more deeply in subsequent work by Soper, who in two seminal papers~\cite{Soper2005Asym,Soper2007} first showed that diffraction data do not contain enough information to discriminate between structure models of strongly different H-bond topology and then that a combination of x-ray (sensitive to O-O and O-H correlations) and neutron diffraction data (sensitive to O-H and H-H correlations) is required to obtain reliable estimates of the three PCFs. Considering the significantly reduced height of the first O-O peak it was concluded that softer MD potentials were called for;~\cite{Soper2007} similar conclusions were reached based on RMC fits to the same data sets.~\cite{leetmaa2008diffraction,Wikfeldt2009} Indeed, actually fitting the Hura {\it et al.} data set using either EPSR~\cite{ Soper2007} or RMC~\cite{leetmaa2008diffraction,Wikfeldt2009} gives a first peak height (2.3) and position (2.82-2.85 {\AA}) in agreement with the analysis by Narten and coworkers~\cite{Narten1971,Narten1970} of their earlier data as well as with the Fourier transforms of recent more extended data sets.~\cite{Fu2009,NeufeindMolPhys,Huang2011}

\begin{figure}[h] 
\includegraphics[angle=0,scale=0.42]{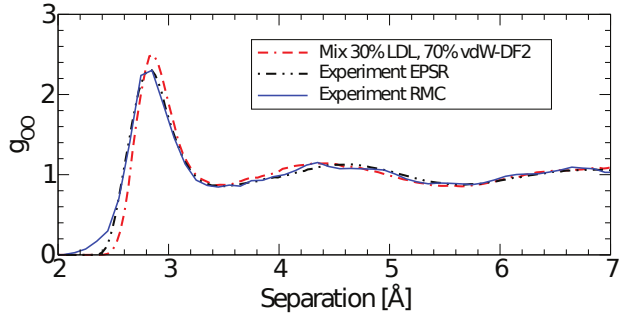}
\caption{Mixing of experimental LDL and vdW-DF2 oxygen-oxygen PCFs in comparison with PCFs from reverse Monte Carlo (RMC) ~\cite{Wikfeldt2009} and EPSR analyses (EPSR) of experimental data.~\cite{Soper2007}}
\label{mix}
\end{figure}

We now test whether the obtained O-O PCF from the vdW models, with their low and asymmetric first peak at long distance and smeared out second shell, can be compatible with the PCF for ambient water under the assumption that the interactions are sufficiently well described, but that the simulation protocol may have introduced too strong constraints on possible structures. That HDL-like water would dominate the liquid under ambient conditions, {\it i.e.} the structure found with the vdW fuctionals, would be in agreement with what has been suggested from x-ray spectroscopy, as well as obtained from all scenarios for supercooled water.~\cite{Wernet2004, tokushima2008high, PNAS2009, KumarPNAS-SE, XuNphys-SE} In those scenarios fluctuations between HDL and LDL forms are assumed and, in view of the vdW functionals seemingly giving only HDL-like solutions, we explore whether adding a ''missing'' LDL contribution, as postulated in these scenarios, could give consistency with experiment in terms of the O-O PCF. We thus weigh together the vdW-DF2 O-O PCF with a model of LDL to a combined PCF and compare with the PCF derived from experiment using EPSR~\cite{Soper2007} and RMC.~\cite{Wikfeldt2009} Since the PBE simulated structure is far from its preferred density,~\cite{Wang2010} it can be assumed to have too large distortions from the "real" LDL that could appear as fluctuations in the otherwise HDL dominated liquid, and we thus compare with the experimental LDL PCF from Soper and Ricci.~\cite{soper2000highdens} 

In fitting to the experimental O-O PCF we obtain agreement (Fig. \ref{mix}) with a 70:30 mixture between vdW-DF2 HDL PCF and the experimentally derived LDL PCF.~\cite{soper2000highdens} This ratio is most interesting, since it is very close to the original estimate of Wernet {\it et al.}~\cite{Wernet2004} and the estimation based on x-ray emission spectroscopy,~\cite{tokushima2008high,PNAS2009} as well as to that from interpreting infrared data in connection with analysis of a fractional Stokes-Einstein relation in water.~\cite{XuNphys-SE} Note furthermore, that quantum effects have not been included in the simulations which would be expected to bring down and broaden the first O-O correlation additionally.~\cite{ Fanourgakis2006,Paesani2007, Walker2010}

As has been pointed out by Soper~\cite{Soper2010} when combining two separate PCFs one must also consider whether the combination introduces additional cross-terms between the two, {\it i.e.} that the contribution to the total PCF from considering pairs of atoms, one from each distribution, could change the picture. This would be expected from a combination of two highly structured PCFs with well-defined peaks occurring at different interparticle separations in the two distributions. However, considering that both the LDL and HDL local structures give a peak in the region of 2.7~-~3 {\AA} and beyond that the HDL-like PCF is basically without structure it seems likely that in this particular case no extra features should be expected from cross contributions to a combined PCF. 

The question is naturally why the vdW simulation only shows the appearance of HDL-like water and why, in order to obtain agreement with x-ray diffraction experiments, it is necessary to artificially add an LDL component. The fact that a combination of an experimental LDL O-O PCF and that from vdW quite accurately reproduces the latest O-O PCF of ambient water is of course no proof that real water is a combination of the two. However, the increased accuracy of the interaction potential obtained with these latest generation vdW functionals indicates that other causes than the non-local interaction should be explored to account for the discrepancy between simulated and measured PCF. 

One potential explanation could be related to the fact that the simulation is performed in the NVE ensemble, which keeps the volume fixed and thus does not allow fluctuations of the density of the box and that this penalizes LDL to a greater extent than HDL, once the more isotropic vdW interactions are included; the NVE ensemble is equivalent to adding a pressure to maintain the box size which would disfavor fluctuations to lower density assuming that the density at ambient conditions corresponds more closely to that of HDL. The box is furthermore rather limited with only 64 molecules. In order for spatially separated fluctuations between HDL and LDL to develop fully it might be necessary to use much larger simulation boxes, in particular if the fluctuations are of a mean length scale around 1 nm as suggested in Refs. \citenum{PNAS2009} and \citenum{Huang2011}. There is furthermore some experimental evidence from thin water films on slightly hydrophobic surfaces that only an HDL related structure is observed even in the supercooled regime,~\cite{Kaya-BaF2} indicating that if the system size becomes very small, indeed only one class of local structure is observed and the formation of LDL-like local regions is suppressed.

\section{Conclusions}

The new van der Waals density functionals optPBE-vdW and vdW-DF2 show great promise in describing the basic structural constituents of liquid water, as seen from comparing calculations of water dimer and hexamers with benchmark coupled cluster CCSD(T) results.~\cite{Kelkkanen2009,Klimes2010,Lee2010} A softening of the structure of liquid water at ambient conditions is observed when including vdW interactions, consistent with previous work.~\cite{Lin2009, Schmidt2009,Wang2010} This is seen from the broader angular distributions, the more disordered tetrahedrality and asphericity distributions, and from the much lower and broader first peak of the oxygen-oxygen PCF obtained from the optPBE-vdW and vdW-DF2 models compared to PBE. The lower first peak of the O-O PCF improves the agreement with experiment significantly. However, the outer structure is washed out by the vdW forces. This has been suggested~\cite{Wang2010} to be related to non-local correlations, but our study of functionals with different non-local correlation strength did not show any significant difference in the liquid structures, while both were found to be very accurate for the water dimer. Instead we find that the inclusion of the more isotropic vdW interaction shifts the balance over from directional H-bonding towards a more close-packed system, {\it i.e.} a competition between directional and isotropic interactions.

The vdW simulations seem to be potentially consistent with a picture of fluctuations between two different water structures instantaneously coexisting in nanoscale patches albeit not directly observing fluctuations except in the sense of obtaining two alternative endpoints with vdW forces included (HDL) or excluded (LDL). The relatively small simulation can only give a picture of the local structure of water, and while PBE predominantly describes an approximation to low-density water, both optPBE-vdW and vdW-DF2, as well as vdW-DF~\cite{Wang2010}, describe an approximation to high-density water. By comparing the O-O PCFs of the vdW models with PCFs from x-ray~\cite{Okhulkov1994} and neutron~\cite{soper2000highdens} diffraction of water at different pressures we note a resemblance between the vdW models and high-density water in terms of effects on the second- and third-neighbor correlations while the expansion of the first coordination sphere found in the simulations may in experiments be counteracted by the pressure applied to experimentally generate pure HDL. The comparison to HDL is further supported by the reduction of the average number of H-bonds per molecule in the vdW MD simulations, which is a result of the isotropic vdW forces competing with the directional H-bond formation. Varying the strength of the exchange interaction does not result in a significant change in number of bonds once the vdW interaction is included. A 70:30 mixture of vdW-DF2 and the experimentally determined LDL PCF is compatible with the latest x-ray O-O PCF which, however, does not constitute proof of a fluctuating real water structure, but indicates the possibility that averaging over a trajectory obeying less restrictive simulation conditions in terms of box size, length of trajectory etc. could result in an O-O PCF directly comparable with experiment.

Quantum effects are not included in the current simulations but including them should not qualitatively change the consistency with the presented picture. The internal R$_{OH}$ bond distance is kept fixed during the simulations which might affect the hydrogen bonding, but not the comparison between PBE and the vdW functionals. Lastly, the possibility that the vdW interaction is not completely accounted for by the current vdW functionals still exists although calibration against various benchmarks indicate a quite reliable representation.

The present work does not resolve the debate on water structure but it suggests for further investigation the van der Waals interaction as a physically sound mechanism which affects the balance between directional H-bonding and higher packing and may thus indicate a way to reconcile the interpretation of recent x-ray spectroscopic data with structures obtained from AIMD simulations of liquid water. It is likely that much larger and longer simulations in the NPT ensemble are needed to determine whether current vdW models support a temperature-dependent balance of fluctuations between HDL and LDL-like structures in ambient water, as suggested by recent x-ray spectroscopic and diffraction results,~\cite{PNAS2009} and which would be enhanced upon cooling, as they must according to all scenarios for water at supercooled temperatures. From the present work it is, however, clear that a consistent description of the vdW interaction in AIMD simulations may possibly provide the key to tuning such a balance.

\section{Acknowledgement}

The authors would like to thank the Lundbeck foundation for sponsoring the Center for Atomic-scale Materials Design (CAMD) and the Danish Center for Scientific Computing for providing computational resources. The authors would also like to thank the Department of Energy, Basic Energy Sciences, National Science Foundation (CHE-0809324), Denmark-America Foundation and the Swedish Research Council.


\end{document}